\documentclass{article}

\usepackage{authblk}
\usepackage[utf8]{inputenc}
\usepackage[left=2cm,right=2cm,top=1.5cm,bottom=1.5cm]{geometry}
\usepackage{amsmath}
\usepackage{amssymb}
\usepackage{graphicx}
\usepackage{subfig}
\usepackage{mathtools}
\usepackage{mathpazo}
\usepackage[mathpazo]{flexisym}
\usepackage{breqn}
\usepackage{verbatim}
\usepackage{xcolor}
\usepackage{ulem}
\usepackage{placeins}
\usepackage{hyperref}
\usepackage{soul}
\hypersetup{
    colorlinks=true,
    linkcolor=blue,
    filecolor=magenta,      
    urlcolor=cyan,
    citecolor=teal
    }
\usepackage{comment}
\graphicspath{{figs/}}

\usepackage[style=numeric-comp,style=phys,url=true]{biblatex}
\DeclareMathOperator{\sech}{sech}
\DeclareMathOperator{\csch}{csch}

\title{Parametric Resonance in the Non-Autonomous Sine-Gordon Model}
\author[1]{Tomasz Dobrowolski}
\author[2]{Jacek Gatlik}
\author[2]{Zofia Bryłowska}
\author[3]{Panayotis G. Kevrekidis}
\affil[1]{\textit{Department of Physics and Applied Mathematics, University of the National Education Commission in Krakow, Podchor\c{a}\.zych  2, 30-084 Cracow, Poland}}
\affil[2]{\textit{AGH University of Krakow, Faculty of Physics and Applied Computer Science, 30-059 Krakow, Poland}}
\affil[3]{\textit{Department of Mathematics and Statistics, University of Massachusetts, Amherst, Massachusetts 01003-4515, USA}}
\date{\today}

\begin{document}
\maketitle

\begin{abstract}
The sine-Gordon model is studied with model parameters that depend on both space and time.
An effective model with one degree of freedom is constructed, allowing the description of the kink movement in both a temporally 
non-autonomous and spatially inhomogeneous setting.
As a highly demanding test of the reduced  model, the case of a temporal drive leading
to a parametric instability is considered. Here, 
the boundaries of the instability region in the form of Arnold tongues were examined in both the simplified effective model  and the full field model. Good agreement was obtained as concerns the regions of stability. It was observed 
that only {in the bottom parts}  of Arnold's tongues 
is the dynamics of the field model more complicated than in the approximate model.
\end{abstract} \hspace{10pt}

\section{Introduction}
Solitons are remarkably stable, localized wave structures that maintain their shape while propagating at constant speed, arising from a delicate interplay between nonlinearity and dispersion in the medium \cite{Drazin1989, Remoissenet1999}. Originally identified in shallow water waves \cite{Russell1844}, solitons have since been identified in a wide variety of physical, chemical, and biological systems, providing deep insights into natural phenomena and driving numerous technological advancements. In polymer physics, solitons contribute to energy transport along polymer chains and offer a microscopic framework for understanding structural defects, their mobility, and the associated transitions \cite{Heeger1988}. This understanding is vital for designing materials with tailored mechanical and electrical properties. In biological systems, solitons have been proposed to mediate low-frequency collective excitations in proteins and DNA, facilitating energy transfer and conformational changes essential for enzymatic activity and genetic regulation \cite{Davydov1977,Lakhno2009,Ivancevic2013}. Their existence underscores the role of nonlinear excitations in fundamental biological processes.

Among the models of relevance to physical systems, the sine-Gordon equation plays a pivotal role in describing phenomena such as, e.g., the Kosterlitz-Thouless transition in two-dimensional systems, as captured by the classical XY model \cite{Kosterlitz1973}. In superconducting electronics, the sine-Gordon equation governs the dynamics of gauge-invariant phase differences in Josephson junctions, revealing soliton behaviour of potential relevance to quantum computing and superconducting circuits \cite{Ustinov1998}. 
Solitons, frequently modeled by the sine-Gordon equation, in systems such as coupled torsional pendula and surfaces of constant negative curvature continue to occupy a central position in scientific inquiry and technological innovation across a wide range of disciplines~\cite{CKW14}. Indeed, the relevant model continues
 to emerge in novel areas including,  most recently, different variants
 of cold atoms and Bose-Einstein condensates~\cite{sotiriadis,siovitz2024doublesinegordonclassuniversal}.

Lately, the study of far-from-equilibrium dynamics in physical systems has gained considerable momentum, particularly in condensed matter settings. Indeed, this growing focus is largely fueled by breakthroughs in cold-atom experiments, which offer highly controllable environments for probing non-equilibrium phenomena. By applying targeted perturbations, such systems can be driven far from equilibrium, posing deep challenges in understanding their long-time behaviour. One prominent area of investigation involves parametric resonance in atomic Bose-Einstein condensates (BECs), which has been the subject of sustained theoretical \cite{Nicolin2007,Ripoll1999,Tozzo2005,Kramer2005} and experimental \cite{Modugno2006,Schori2004,engels,choi} research spanning multiple decades. The feasibility of experimentally realizing these effects arises from the relative simplicity of generating periodic driving, for instance, through modulation of magnetic trap parameters or via periodic tuning of interaction strength using Feshbach resonances \cite{chin}. The latter is a topic of intense current
interest, as evidenced by numerous recent works~\cite{choi,liebster2025observationsupersolidlikesoundmodes,wang2025parametricexcitationsharmonicallytrapped}.

Another, more elementary example of a system in which parametric resonance occurs is the Kapitza pendulum \cite{Kapitza1951} which is a good example of a system that is subject to periodic parameter changes. This system is a rigid pendulum whose pivot point is subject to a harmonic drive. It turns out that the sine-Gordon model containing a periodic time dependence, via the strength of the cosine potential, can be understood as the continuous limit of a system of multiple coupled such oscillators. The dynamical stability of periodically-driven sine–Gordon model has been the subject of studies in the area of finite amplitudes and frequencies \cite{Citro2015}.

In this study, we explore an extension of the sine-Gordon model that incorporates two coordinate-dependent functions. The physical interpretation of these functions varies depending on the specific application: one possibility involves accounting for geometric curvature effects in Josephson junctions \cite{Gatlik2021}, while another can represent spatial variations in the junction’s thickness \cite{McLaughlin1978}. Our focus, however, lies on a broader generalization in which the coefficient in front of the potential term is allowed to vary explicitly in time.
The primary objective is to construct a reduced, effective model involving a single collective coordinate that can faithfully reproduce the dynamics of the kink field configuration, 
especially in regimes that deviate significantly from those described by the standard sine-Gordon theory. 
As implied above, we concentrate specifically on the topological sector characterized by kink-like excitations.
Furthermore, we investigate how external periodic driving affects the kink dynamics. In particular, we consider a traveling perturbation with fixed wavevector and frequency. This form of excitation induces rich and nontrivial dynamical phenomena, rendering the system highly responsive 
to (and, indeed, potentially resonant with) the spatiotemporal modulation.

The structure of the article is as follows. Section 2 introduces the model under investigation. In this part, we describe the functions used to deform the sine-Gordon model and outline the initial and boundary conditions employed in the numerical simulations. In section 3, we derive the effective model.
Section 4 contains, on the one hand, a comparison between the trajectories obtained from the effective model and those computed within the original field-theoretic framework, and on the other hand, a stability analysis based on a linear approximation in a small parameter present in the definition of the system.
Our results are summarized in the concluding section 5.

\section{System description}
Here, we focus on the version of the sine-Gordon model, expressed as a partial differential equation (PDE) given by:
\begin{equation}
\label{sine-gordon}
\partial_t^2\phi  - \partial_x(\mathcal{F}(x)\partial_x\phi) + g(t,x) \sin(\phi)  = 0,
\end{equation}
where
\begin{equation}
\label{f_function}
\mathcal{F}(x) = 1 + \varepsilon_2\sin\left(\frac{ \pi }{12} x\right) \mathrm{ ~~~~ and ~~~~ } g(t,x) =1+\varepsilon_1\sin(k x -\omega t).
\end{equation}
 The $\mathcal{F}(x)$ function appears, for example, in the context of Josephson junctions, where its role is to take into account effects related to curvature \cite{Dobrowolski2012,Dobrowolski2009,Gatlik2021}.
In the same context, the modulation function $g(t,x)$ may represent changes in Josephson current or the presence of 
external alternating electromagnetic field, which periodically enhances or suppresses the Josephson response, associated
with the phase field $\phi$ by modulating the amplitude of the critical current. 
In this study, we introduce a mechanism through which changes in the critical current can trigger fluxon movement.
It is important to emphasize that the function $g$ depends not only on spatial coordinates but also varies with time. Specifically, its time dependence manifests as a traveling wave characterized by wave vector $k$ and frequency $\omega$. Under appropriate parameter regimes, such a perturbation can enable kink transport even in the absence of external forcing. The initial conditions for the system are chosen as follows:
\begin{equation}\label{phi_wp1}
\phi(0,x)=4 \arctan \left[ \exp \left( \sqrt{\frac{g(t=0,x_0)}{{\mathcal{F}(x_0)}}}  \, (x-x_0) \right) \right],
\end{equation}
\begin{equation}\label{phi_wp2}
\partial_t \phi(0,x)= - 2 v \sqrt{ \frac{g(t=0,x_0)}{\mathcal{F}(x_0)}}\sech \left[  \sqrt{\frac{g(t=0,x_0)}{\mathcal{F}(x_0)}}  \, (x-x_0) \right].
\end{equation} 
This unusual choice of initial conditions is related to the ansatz proposed in this work. Moreover, this choice means that the configuration determined by these initial conditions in the situation under consideration carries energy almost equal to the lowest energy of the kink in the considered configuration,  i.e., for given functions ${\cal F}$ and $g$, a given position and velocity of the soliton.
This configuration enables us to prevent both the excitation and the resulting energy radiation from the field profile. Moreover, we impose standard kink-like Dirichlet boundary conditions that are consistent with the chosen initial conditions.

\section{Effective Model}
The construction of the effective description for the system described by the equation \eqref{sine-gordon}  is based on the Lagrangian
\begin{equation}\label{L}
L= \int_{-\infty}^{+\infty} dx {\cal L}(\phi)
=\int_{-\infty}^{+\infty} dx  \left[ \frac{1}{2}~ (\partial_t
\phi)^2 - \frac{1}{2} ~{\cal F}(x) (\partial_x \phi)^2 - g(t,x)
(1-\cos \phi) \right].
\end{equation}
An effective description of the kink dynamics can be obtained by switching from the original field variable $\phi(t, x)$ to a new one, $\xi(t, x)$, through the transformation
\begin{equation}\label{phi}
\phi(t,x)=4 \arctan e^{\xi(t,x)}.
\end{equation}
After substituting the field $\phi(t,x)$ with $\xi(t,x)$, the Lagrangian takes on the following simplified form:
\begin{equation}\label{Lxi}
L=4 \int_{-\infty}^{+\infty} dx  \,\, {\rm sech}^2 \xi \left[
\frac{1}{2}\,(\partial_{t} \xi)^2 - \frac{1}{2}\, {\cal
F}(x)(\partial_{x} \xi)^2 -\frac{1}{2} ~g(t,x) \right].
\end{equation}
Here, we have applied the identity 
 $1-\cos \phi = 2 {\rm sech}^2 \xi$, which is satisfied by the function specified in Eq. 
~\eqref{phi}.
Now we reduce the field model, i.e. a system with an infinite number of degrees of freedom, to a model containing a single dynamic variable. The price of this reduction is limiting the theory to the sector containing a solution of the single kink type.
Nevertheless, the study of such structures is so 
widespread~\cite{Kivshar1989,CKW14} that this is a 
meaningful restriction.
We perform the above reduction by assuming the following ansatz:
\begin{equation}\label{xi}
 \xi(t,x)= \sqrt{ \frac{g(t,x_0(t))}{{\cal
F}(x_0(t))}} \,\,  (x-x_0(t)).
\end{equation}
The dynamical variable $x_0(t)$ denotes the position of the point at which the field $\phi(t,x)$
reaches the value $\pi$. We identify this point with the location of the kink.
A distinctive feature of the above approach, in contrast to the standard one, is the presence of a term involving both functions $g$ and ${\cal F}$.
Upon performing the integration over the spatial coordinate, one obtains the effective Lagrangian:
\begin{equation}\label{Leff}
L_{eff} = \frac{1}{2}\, M \Dot{x}_0^2  + \alpha \Dot{x}_0
 - V.
\end{equation}
The parameters entering the Lagrangian are defined through the corresponding integrals listed in the appendix. As expected, all coefficients in the Lagrangian depend on the variable $x_0$. Moreover, since the function $g$ depends explicitly on time, the coefficients inherit this time dependence. The resulting equations of motion derived from this Lagrangian are given by:
\begin{equation}
\begin{gathered}
\label{1dof_ansatz}
    M\Ddot{x}_0 +\frac{1}{2}(\partial_{x_0}M)\Dot{x}_0^2+ 
    (\partial_t M) \Dot{x}_0  + \partial_t \alpha +   \partial_{x_0}V=0.
\end{gathered}
\end{equation}
Here, the temporal derivatives capture only the explicit dependence on time, which enters the coefficients through the function $g$.

\FloatBarrier

\section{Kink Dynamics and Resonant Structures in Effective and Field Model}
\begin{figure}[h!]
    \centering
    \includegraphics[height=4cm]{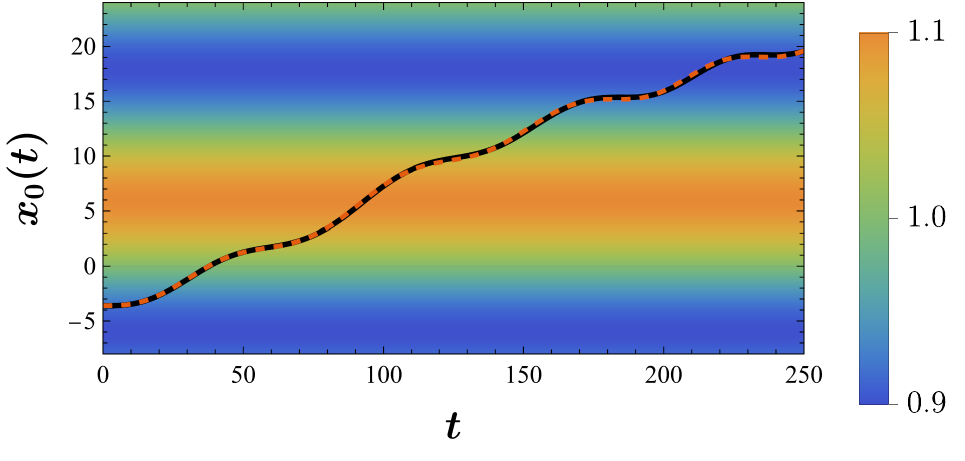}
    \caption{Position of the kink as a function of time, here $\varepsilon_1 = 0.1$, $\varepsilon_2 = 0.1$, $k=\pi/6$, $\omega=0.05$ the initial velocity is $v=0$, and the initial position $x_0=-3.6$. The legend on the right side of the figure shows the values of the function $\mathcal{F}$.}
    \label{fig_B01}
\end{figure}

We begin by comparing the kink's behaviour within a certain parameter range in both the field model and the effective model. Agreement between the effective and full field models is particularly resonable when the parameters 
$\varepsilon_1$  and $\varepsilon_2$ are small.
For example, Figure \ref{fig_B01} shows the course of the trajectories obtained on the basis of the field model \eqref{sine-gordon} (black solid line) and the effective model \eqref{1dof_ansatz} (orange dashed line). Here, both $\varepsilon_1$ and $\varepsilon_2$ have been set to $0.1$. In the simulation, the kink initially rests at $x_0=-3.6$, while the other parameters are $k=\pi/6$ and $\omega=0.05$. The colors in the figure represent the $\mathcal{F}$ function according to the legend on the right. As one can see in the figure, the process of pushing the kink from one minimum to the neighboring one is the result of the analytical form of a $g$ function that is like a running wave (characterized by wave vector $k$ and frequency $\omega$). 
It is worth noting that in the model under consideration, there is no external forcing such as a bias current. Nevertheless, the specific time dependence of the  $g$ function enables the kink to be transported from one minimum of the  ${\cal F}$ function to a neighboring one, overcoming the barrier that separates them. It is worth noting that the agreement between the two models remains excellent  throughout the process, even at times as late as $t=250$ units.

\begin{figure}[h!]
    \centering
    \includegraphics[height=4cm]{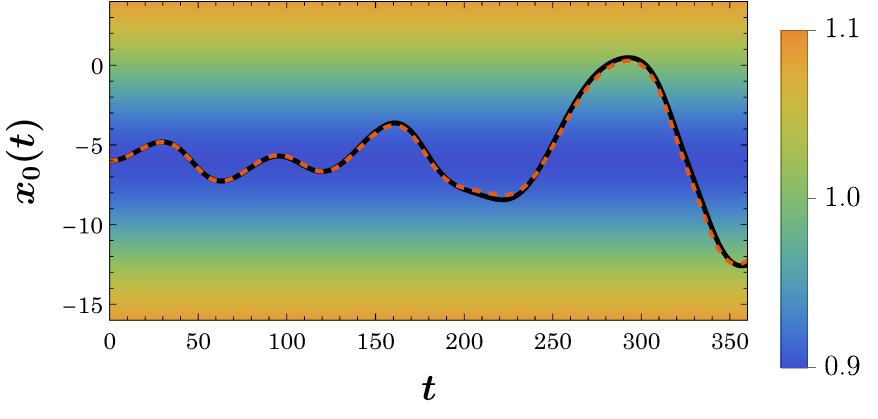}
    \caption{Vibration of the position of the kink around the minimum of function $\mathcal{F}$. Here, $\varepsilon_1 = 0.03$, $\varepsilon_2 = 0.1$, $k=\pi/6$, $\omega=0.1$ the initial velocity is $v=0$, and initial position $x_0=-6$.}
    \label{fig_B02}
\end{figure}

Figure \ref{fig_B02} illustrates the motion of the kink near the minimum of the function ${\cal F}$, also for small values of the parameters $\varepsilon_1$ and $\varepsilon_2$. The non-trivial character of the trajectory arises from an interplay between the natural frequency of small oscillations around the minimum of ${\cal F}$ and the effect of the frequency $\omega$ appearing in the function $g$. The legend in the figure corresponds to values of the function ${\cal F}$. The trajectory obtained from the full field-theoretic model is shown as a solid black line, while the trajectory derived from the effective model is indicated by a dashed orange line. In the simulation, the kink is 
initially placed at $x_0=-6$. The parameters used in the figure are $\varepsilon_1=0.03$, $\varepsilon_2=0.1$, $\omega=0.1$ and $k=\pi/6$. It is apparent that the explicit time dependence of the function $g$ is responsible, under the assumed initial conditions, not only for initiating the kink’s motion, but also for the non-trivial shape of its trajectory.

\begin{figure}[h!]
    \centering
    \includegraphics[height=4cm]{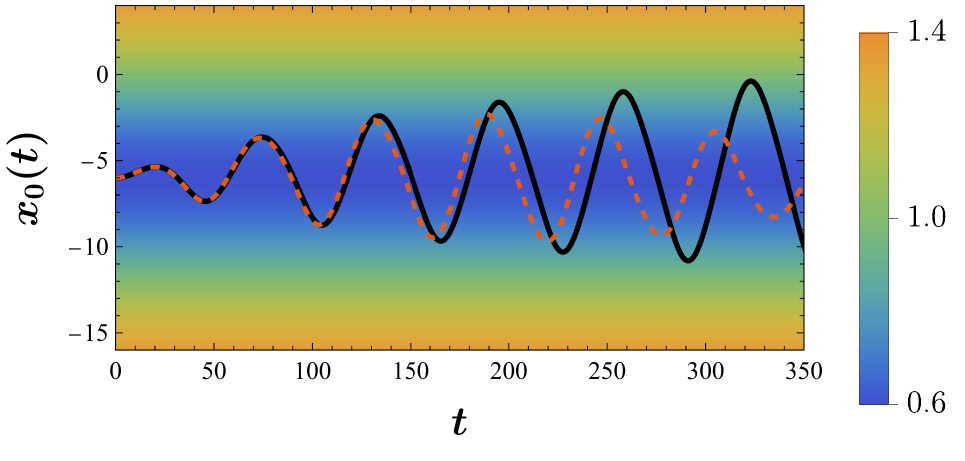}
    \caption{Vibration of the position of the kink around the minimum of function $\mathcal{F}$. Here, $\varepsilon_1 = 0.05$, $\varepsilon_2 = 0.4$, $k=\pi/6$, $\omega=0.1$ the initial velocity is $v=0$, and initial position $x_0=-6$.}
    \label{fig_B03}
\end{figure}

The subsequent figure (Fig. \ref{fig_B03}) shows the vibration process around the minimum of the potential with a considerably larger $\varepsilon_2$. As a reminder, the value of this parameter describes the height of the barrier.
As in the previous figure, the black line represents the trajectory obtained on the basis of the full model \eqref{sine-gordon}, while the orange dashed line is obtained from the evolution of the effective model with one degree of freedom \eqref{1dof_ansatz}. 
It can be seen that in Figure \ref{fig_B03} (for $\varepsilon_1=0.05$) the fit is reasonable up to time $t=200$. Here, the choice of a relatively large parameter value, $\varepsilon_2=0.4$, which is beyond the small parameter regime, noticeably reduces the agreement between the models.

Let us focus on small oscillations of the kink around the minimum of the potential.
First we notice that
the static solution of the above equation can be obtained in the absence of time-dependent parameter, i.e., for $\varepsilon_1=0$, as a solution of the equation
\begin{equation}
\begin{gathered}
\label{xs}
\partial_{x_0}V\mid_{x=x_s}=0.
\end{gathered}
\end{equation}
Naturally, the equation indicates,
the location of the extrema of the potential.
To investigate the soliton oscilations, we focus on a region centered around the potential minimum at $x_s=-6$.
Linearization of the equation \eqref{1dof_ansatz+} around the equilibrium position $x_s$, i.e., $x_0=x_s+\delta x_0$ leads to the equation
\begin{equation}
\begin{gathered}
\label{1dof_ansatz+}
    \delta \Ddot{x}_0+
    \frac{\partial_t M(t,x_s)}{M(t,x_s)} \delta \Dot{x}_0   + \frac{\partial_t \alpha(t,x_s)}{M(t,x_s)}+  \frac{\partial_t \partial_{x_0}\alpha(t,x_0=x_s)}{M(t,x_s)} \delta x_0 +   \frac{\partial^2_{x_0}V(t,x_0=x_s)}{M(t,x_s)} \delta x_0=0.
\end{gathered}
\end{equation}
We can simplify the equation significantly if we assume that  $\varepsilon_1$ is small quantity, then expand the above equation with respect to this parameter and leave only linear terms. The equation obtained in this way has the form of a harmonic oscillator equation with a time-dependent frequency and a dissipation term (thus, naturally suggesting
the possibility of a parametric resonance):
\begin{equation}
\begin{gathered}
\label{Mathieu}
    \delta \Ddot{x}_0
   + \varepsilon_1 \left( b_1 \sin \omega t - b_2 \cos \omega t \right) \delta \Dot{x}_0 +\left( \Omega^2  + \varepsilon_1  a_1  \cos \omega t +\varepsilon_1 a_2  \sin \omega t  \right) \delta x_0=0.
\end{gathered}
\end{equation}
The parameters in the equation are functions of the wave vector $k$ and the frequency $\omega$
\begin{equation}
\begin{gathered}
\label{parameters}
   a_1 = A \sin 6k, ~~~~ 
   a_2 = A \cos 6k,\\
   b_1 = \frac{1}{2} \omega  \sin 6k , ~~~~b_2 =  \frac{1}{2} \omega  \cos 6k ,\\
 \Omega^2 =   - \frac{1}{4} \left(\frac{\pi^2}{12}\right)^2 \frac{\varepsilon_2^2}{1-\varepsilon_2} + \varepsilon_2  \left( \frac{\pi}{12}\right)^4 \left( 3 \sqrt{1-\varepsilon_2} + \varepsilon_2 \left(\frac{\pi}{4} \right)^2 \coth \left(\frac{\pi^2}{24} \sqrt{1-\varepsilon_2} \right) \right) \csch\left(\frac{\pi^2}{24} \sqrt{1-\varepsilon_2} \right)  ,\\
  A = -3 \varepsilon_2 \omega^2 \left( \frac{\pi}{12}\right)^4 +\frac{1}{4} \left(k^2 + \left( \frac{\pi}{12}\right)^2 \right) \varepsilon_2 + \frac{1}{2} k \pi (\sqrt{1-\varepsilon_2})^3 \left(k^2 - \left( \frac{\pi}{12}\right)^2  \frac{\varepsilon_2}{1-\varepsilon_2}\right) \csch \left( \frac{\pi}{2}k \sqrt{1-\varepsilon_2}\right) \\
 + \varepsilon_2 \left( \frac{\pi}{12} \right)^4 \Biggl[ \frac{3}{16} \sqrt{1-\varepsilon_2} \left( 8 + \varepsilon_2 \pi^2 \left( \frac{\pi}{12}\right)^2 \right) - 36 (1-\varepsilon_2) \left(k^2 + \left(\frac{\pi}{12}\right)^2 \right) \coth \left( \frac{\pi}{24} \sqrt{1-\varepsilon_2}\right)\\
 - \frac{1}{24} \left( \frac{\pi}{2}\right)^4 \varepsilon_2 \sqrt{1-\varepsilon_2} \cot \left(\frac{\pi}{24} \sqrt{1-\varepsilon_2} \right) \Biggr] \csch \left( \frac{\pi}{24} \sqrt{1-\varepsilon_2}\right) \\
 -\frac{1}{18} \left( \frac{\pi}{4} \right)^4 k^2 \varepsilon_2 (1 - \varepsilon_2) \csch^3 \left( \frac{\pi}{2} k \sqrt{1-\varepsilon_2} \right) \sinh\left( \pi k \sqrt{1-\varepsilon_2}\right). 
\end{gathered}
\end{equation}
As one can see, the equation itself takes the form of Mathieu’s equation~\cite{magnus2004hill}. In fact, the term containing the first derivative does not describe dissipation here, but the fact that sometimes external forcing is in the opposite phase to the motion of the soliton.
 The term containing the first derivative with respect to the time variable in the above equation can be eliminated by transformation
\begin{equation}
\begin{gathered}
\label{transformation}
    \delta {x}_0(t) = y(t) \exp\left[ \frac{\varepsilon_1}{2 \omega} \left(b_1 \cos \omega t +  b_2 \sin \omega t \right) \right] .
\end{gathered}
\end{equation}
We write the canonical form of Mathieu’s equation using the new variable $y(t)$
\begin{equation}
\begin{gathered}
\label{canonical_M}
       \Ddot{y}
   +\left( \Omega^2  + \varepsilon_1  c_1  \cos \omega t +\varepsilon_1 c_2  \sin \omega t  \right)  y=0.
\end{gathered}
\end{equation}
The constants occurring in this equation are related to the constants used in the equation \eqref{Mathieu} as follows
$c_1=a_1-\frac{1}{2} \omega b_1$ and 
$c_2=a_2-\frac{1}{2} \omega b_2$. Furthermore, in this equation we have omitted the term proportional to $\varepsilon^2_1$, which is inevitably introduced by the transformation \eqref{transformation}.
The stability area boundaries for this equation are determined in the space of parameters $\varepsilon_1, \varepsilon_2$ (or $\Omega$) and $\omega$.

\begin{figure}[h!]
    \centering
    \subfloat{{\includegraphics[height=4cm]{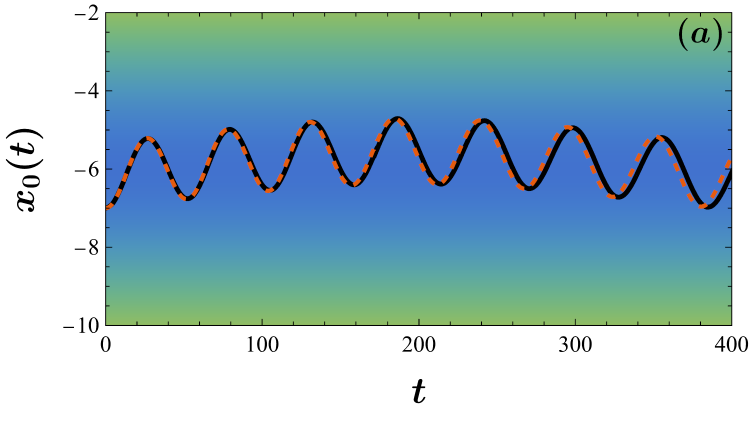}}}
    \quad
    \subfloat{{\includegraphics[height=4cm]{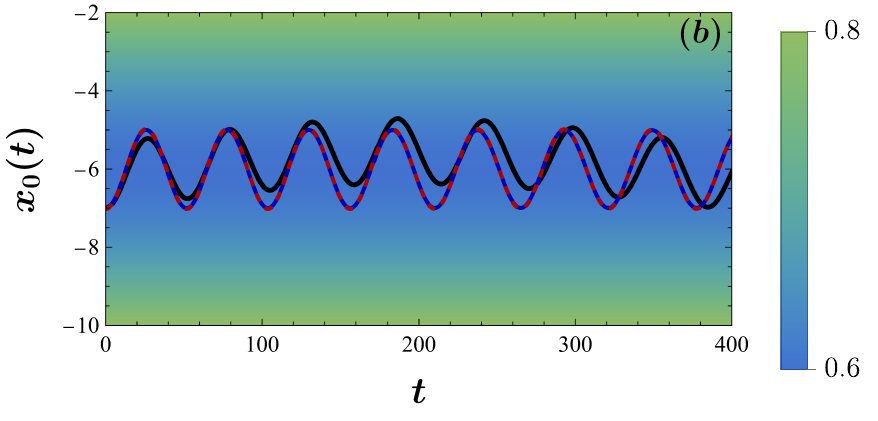}}}
    \caption{Here, $\varepsilon_1 = 0.1$, $\varepsilon_2 = 0.4$, $k=0.2$, $\omega=0.01$ the initial velocity is $v=0$, and $x_0=-7$. Figure (a): The black line corresponds to the full field model solution, while the orange one is obtained on the basis of the effective particle model \eqref{1dof_ansatz}. Figure (b): Comparison of the field model with equation \eqref{Mathieu} (red dashed line) and \eqref{canonical_M} (blue dashed line).}
    \label{fig_B04}
\end{figure}

\begin{figure}[h!]
    \centering
    \subfloat{{\includegraphics[height=4cm]{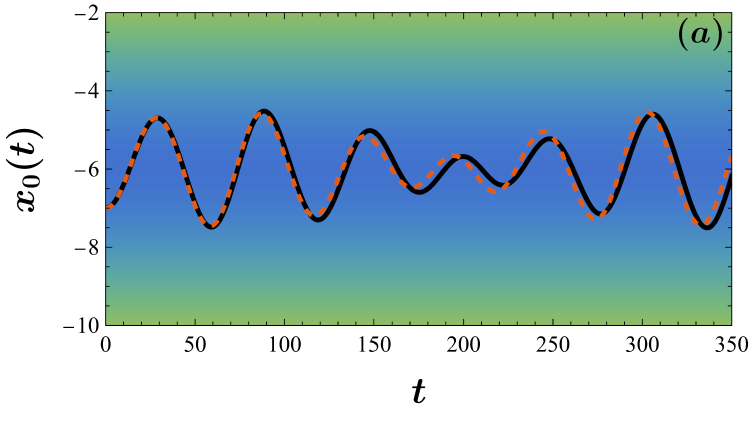}}}
    \quad
    \subfloat{{\includegraphics[height=4cm]{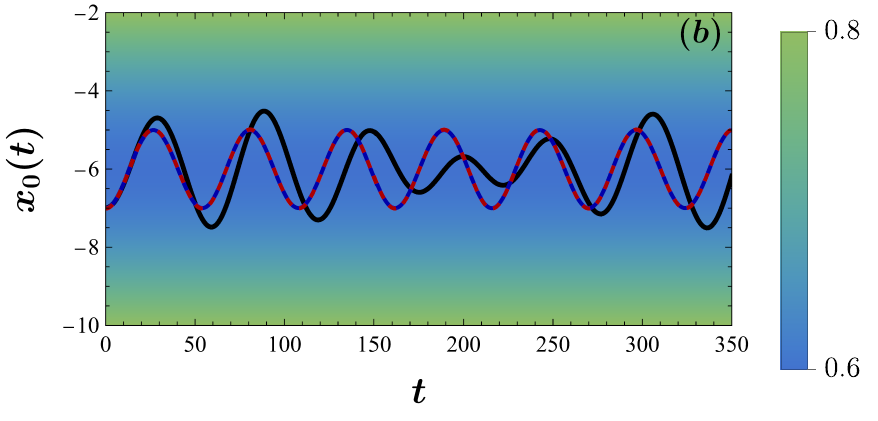}}}
    \caption{Here, $\varepsilon_1 = 0.05$, $\varepsilon_2 = 0.4$, $k=0.2$, $\omega=0.09$ the initial velocity is $v=0$, and $x_0=-7$. Figure (a): The black line corresponds to the full model solution, the orange one is obtained from the model \eqref{1dof_ansatz}. Figure (b): Comparison of the field model with equation \eqref{Mathieu} (red dashed line) and \eqref{canonical_M} (blue dashed line).}
    \label{fig_B05}
\end{figure}

The predictions obtained on the basis of the field model \eqref{sine-gordon} (black solid line) and the one-degree-of-freedom model \eqref{1dof_ansatz} (orange dashed line) are shown in Figure \ref{fig_B04} (left panel).
The initial position of the kink is $x_0=-7$ where the kink initially has speed $v \equiv \dot{x}_0(0)=0$. 
Other parameters are equal to $\varepsilon_1 = 0.1, \varepsilon_2 = 0.4, k = 0.2, \omega = 0.01$. 
{As mentioned above,  the linearized effective equation \eqref{Mathieu} and its form \eqref{canonical_M} differ by terms quadratic in the parameter $\varepsilon_1$. However, it turns out that within the considered parameter range, these differences are negligible. This is illustrated in Figure \ref{fig_B04} b. It can be seen that the agreement/disagreement of both linearized models with the field model is identical at the scale of the plot.}
As one can see, the two models yield very proximal results, up to times exceeding $300$ units. {Due to this convergence and for computational convenience during the stability analysis, we will use equation \eqref{canonical_M}, which is the canonical form of the Mathieu equation.}
For the same initial conditions but for parameters $\varepsilon_1 = 0.05, \varepsilon_2 = 0.4, k = \frac{\pi}{6}, \omega = 0.05$, the trajectory course is more complex but, nevertheless, the
trajectory {obtained from the effective model \eqref{1dof_ansatz}} stays close to the PDE kink center motion up to
about $200$ units (Fig. \ref{fig_B05} left panel). The right {panel of this figure}, compare the trajectory obtained from the field model \eqref{sine-gordon} and two linearized equations \eqref{Mathieu} (blue dashed line) and \eqref{canonical_M} (red dashed line). Note that omitting the quadratic terms in $\varepsilon_1$ when going from equation \eqref{Mathieu} to equation \eqref{canonical_M} has no apparent effect on the trajectory. 
On the other hand, the difference between the field model \eqref{sine-gordon} and equation \eqref{canonical_M} can   occasionally
 be more significant as in Figure \ref{fig_B05} panel (b). 
In the latter case, the maximum deflections are different and the vibration periods for both trajectories are slightly different. The compatibility of linearized models is limited only to times
up to around $t=50$ in this case. {The discrepancy arises from the fact that the linearized approximation equation \eqref{canonical_M} is significantly less sensitive to variations in the driving frequency $\omega$ than either the field equation or the full approximate equation \eqref{1dof_ansatz}.}

Equation \eqref{canonical_M} for $\varepsilon_1=0$ has the form of a harmonic oscillator equation with natural frequency $\Omega$. When we switch on a time-dependent external disturbance, some instability may appear in the system. In our case, this corresponds to the non-zero value of the parameter $\varepsilon_1$. The boundaries of the stability regions are determined, on the one hand, in the parameter plane $(\Omega,\varepsilon_1)$, i.e., in the plane describing the dependence of the amplitude of the forcing 
versus the natural frequency of the undisturbed system. On the other hand, the relevant description can be represented on the $(\omega,\varepsilon_1)$ plane, i.e., on the plane describing the dependence of the $g$ function amplitude change versus  the  frequency of the disturbances.

\begin{figure}[h!]
    \centering
    \subfloat{{\includegraphics[height=4cm]{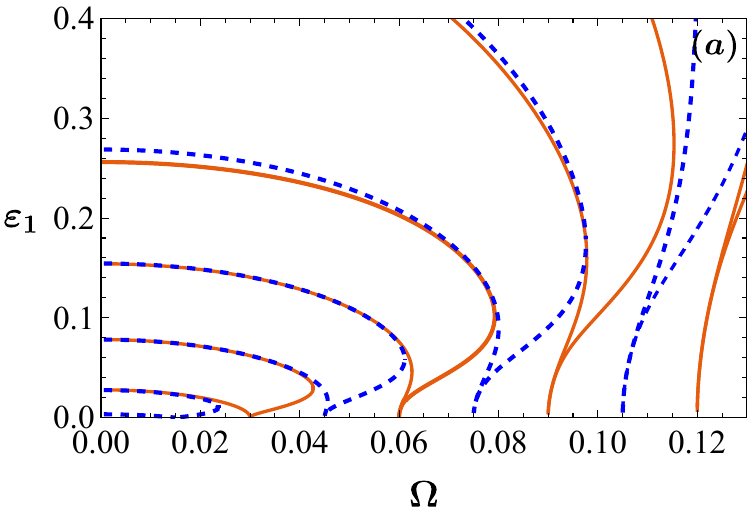}}}
    \quad
    \subfloat{{\includegraphics[height=4cm]{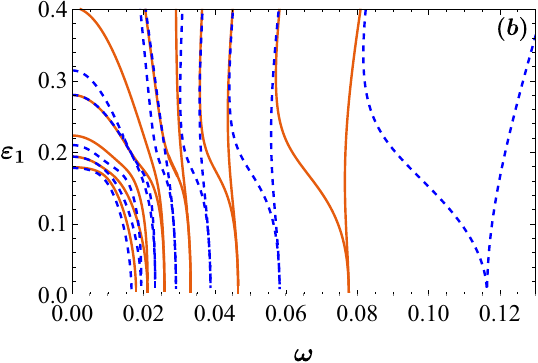}}}
    \caption{Arnold's tongues corresponding to the period $T=\frac{2\pi}{\omega}$ (blue dashed line) and  for period $T=\frac{4\pi}{\omega}$ (orange line), in both figures $k=\frac{\pi}{6}$. The remaining fixed parameters take values on both panels, respectively (a) $\omega=0.03$ and (b) $\varepsilon_2=0.4$.}
    \label{fig_B06}
\end{figure}

Figure \ref{fig_B06} shows the
Arnold's tongues obtained on the basis of equation \eqref{canonical_M}. 
Stable areas are characterized by limited oscillations, while unstable areas, which are located inside the Arnold's tongues, correspond to exponentially increasing solutions.
Unstable regions are plotted for trajectories with period $T=\frac{2 \pi}{\omega}$ (solid orange lines) and for doubled period (dashed blue lines). Doubling the period is a common feature for systems described by Mathieu’s equation; see, e.g., \cite{Ward2025}.  In the figures, wave vector  is set to $k=\frac{\pi}{6}$. In figure \ref{fig_B06} (a), the frequency is $\omega=0.03$, while in figure \ref{fig_B06} (b), the $\varepsilon_2$ parameter is equal to 0.4.

\begin{figure}[h!]
    \centering
    \subfloat{{\includegraphics[height=4cm]{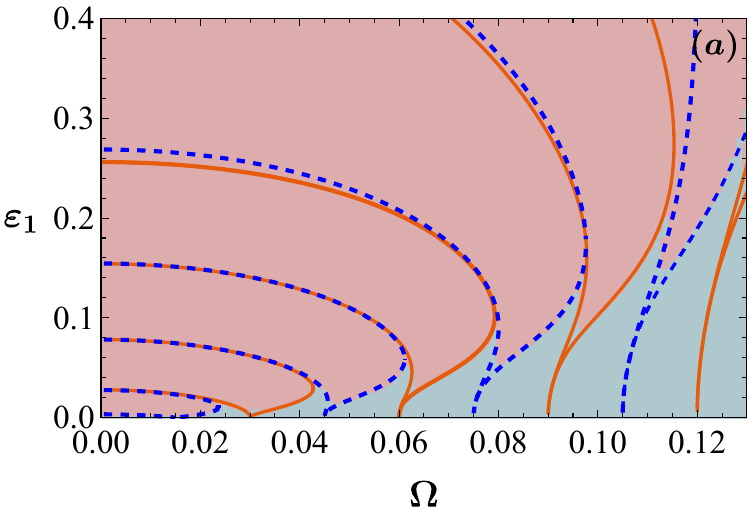}}}
    \quad
    \subfloat{{\includegraphics[height=4cm]{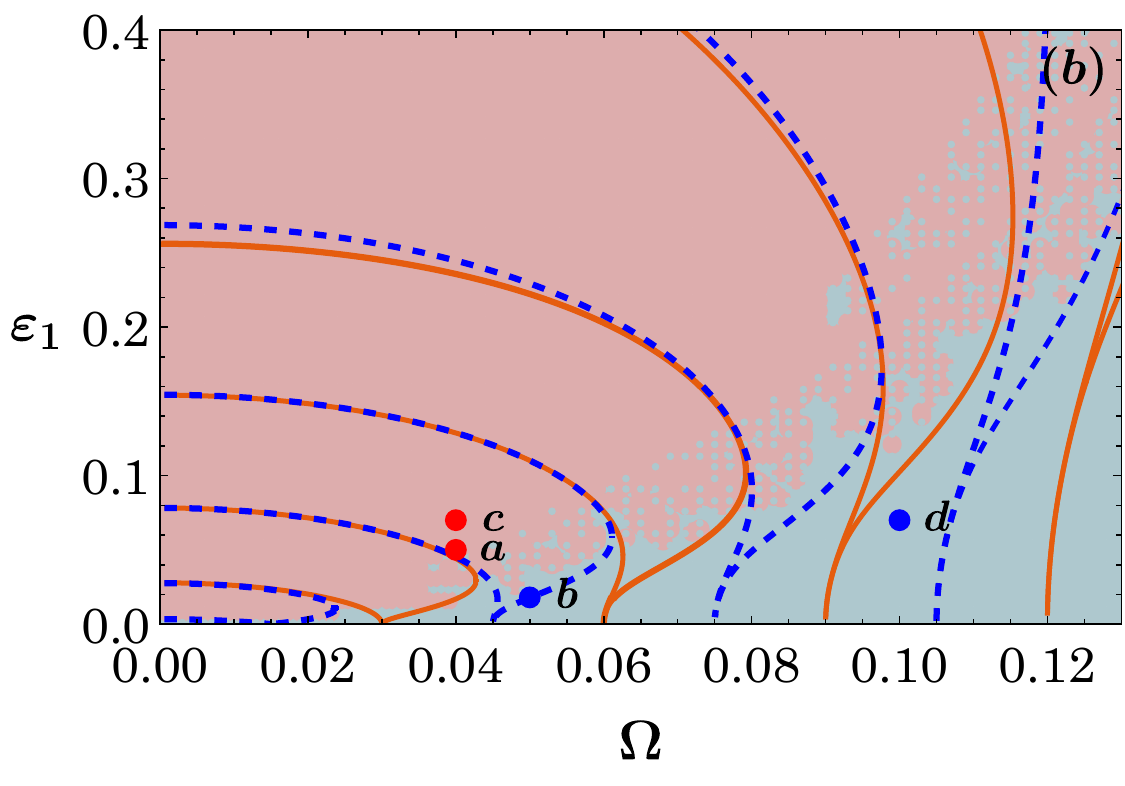}}}
    \caption{Arnold's tongues for 
 parameters $k=\frac{\pi}{6}$ and $\omega=0.03$. (a) The red and blue dashed lines were obtained based on the equation \eqref{canonical_M}, while the light red and blue areas were obtained based on the Mathieu equation \eqref{Mathieu}. (b)  Light red and blue areas obtained based on the field model \eqref{sine-gordon}.}
    \label{fig_B07}
\end{figure}

Figure \ref{fig_B07} (a) contains a detailed comparison of the results shown in Figure \ref{fig_B06} (a) with the results obtained based on equation \eqref{Mathieu}. In this figure, the areas marked in light red show the inside of Arnold's tongues based on the model of Eq.~\eqref{Mathieu} , whereas the red and dashed blue lines were obtained from the equation \eqref{canonical_M}.
It should be noted that on the outside of Arnold's tongues (blue area in the figure), we are always dealing with stable oscillations around the equilibrium position. 
On the other hand, for parameters corresponding to the inside of Arnold's tongues, the escape of kink outside the area where we were dealing with vibrations around the equilibrium position was always observed. 
It can be seen that predictions of both equations  \eqref{Mathieu} and \eqref{canonical_M} are consistent, i.e., neglecting the quadratic terms does not lead to visible effects.

Figure  \ref{fig_B07} (b) shows the areas of instability (marked light red) obtained on the basis of the field model \eqref{sine-gordon} with the results obtained on the basis of equation \eqref{canonical_M} (red and blue dotted lines). The stability regions obtained from the field equation are colored blue. 
{It can be seen that the regions identified as stability regions  based on the Mathieu equation \eqref{canonical_M} also correspond to stability regions determined from the field equation. The situation is somewhat different in the parametric
regions that the Mathieu equation \eqref{canonical_M} predicts to be unstable. In the upper parts of these regions, both equations \eqref{sine-gordon} and \eqref{canonical_M}   exhibit instability (blue region). However, in the lower parts of the tongues, the field equation reveals the existence of some irregular stability regions, even though the Mathieu equation predicts instability. It is worth emphasizing here that in the case of the problem studied in this work, equation \eqref{canonical_M}  is only an approximate equation. 
Indeed, the existence of an infinity of additional degrees of
freedom at the PDE level may occasionally alter the 
stability conclusions (e.g., by rapidly saturating the instability).
In this spirit, we have found that the reduced equation(s)
derived herein correctly indicates the stability regions below the Arnold tongues. The deviations observed within the tongues lie outside the range of validity of the linear approximation, hence these deviations from instability observed in the field simulations do not
invalidate the developed theory; rather they delineate the limits
of its applicability.}
To summarize, the results obtained based on the field equation show that in the upper parts of the tongues, we are always dealing with departing from the stable oscillation region. The situation is different in the lower part of the tongues, where stable vibration and escape areas {alternate in the way shown in Fig.~\ref{fig_B07}.}
Naturally, when it comes to stable vibration inside the tongue areas, our prediction is limited by the simulation times, which were set to $t=3000$. On the other hand the density of points for which simulations were carried out in the field model in both directions (for both axes) of the described figure is at the level of $0.001$.

\begin{figure}[h!]
    \centering
    \subfloat{{\includegraphics[height=4cm]{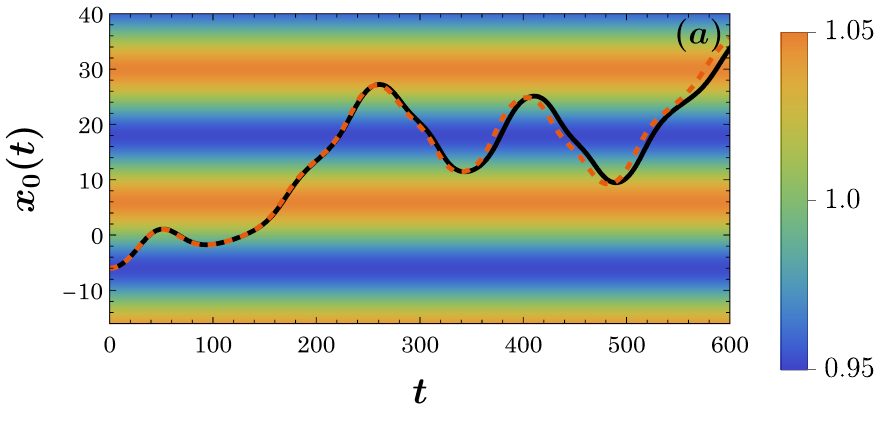}}}
    \quad
    \subfloat{{\includegraphics[height=4cm]{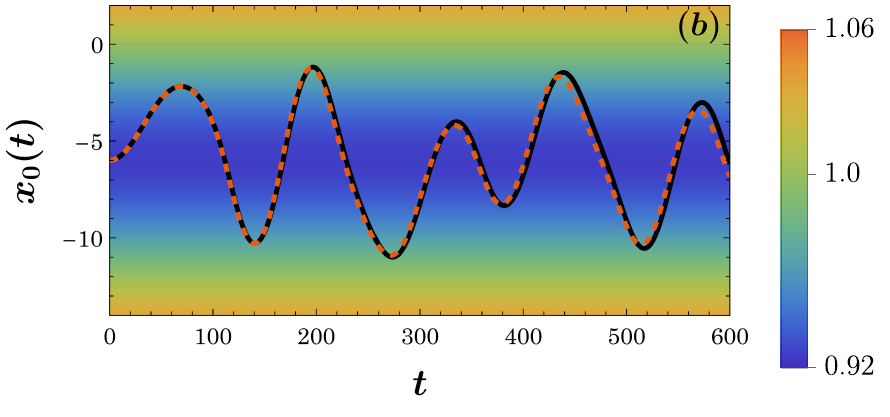}}}
    \quad
    \subfloat{{\includegraphics[height=4cm]{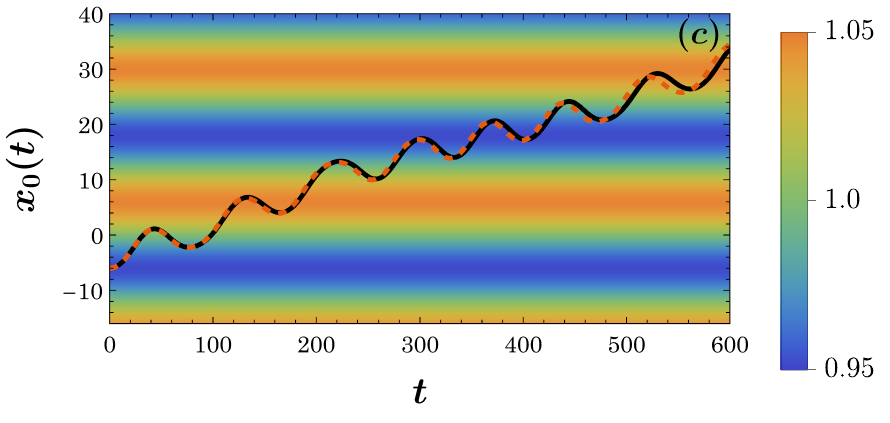}}}
    \quad
    \subfloat{{\includegraphics[height=4cm]{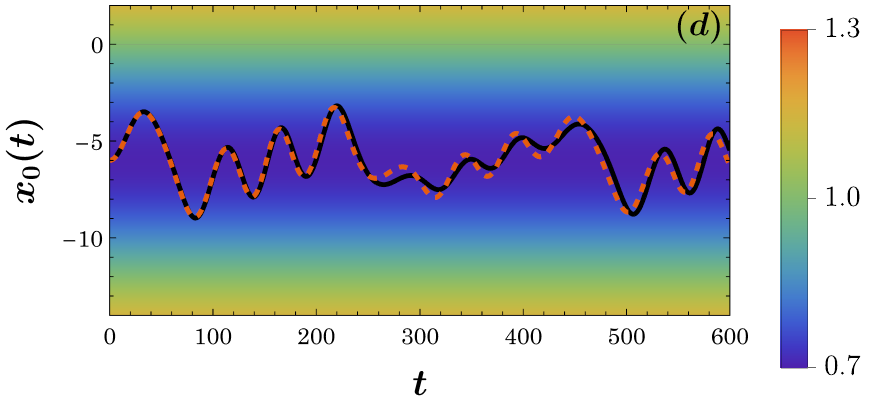}}}
    \caption{Representative trajectories corresponding to the stable and unstable regions from the previous figure. The parameters were selected as follows: $k=\pi/6$,  $x_0(0)=-6$, $\omega = 0.03$ (a) $\Omega=0.04$,  $\varepsilon_1 = 0.05$, (b) $\Omega=0.05$,  $\varepsilon_1 = 0.02$, (c) $\Omega=0.04$,  $\varepsilon_1 = 0.07$ and (d) $\Omega=0.1$,  $\varepsilon_1 = 0.07$.
    }
    \label{fig_B08}
\end{figure}

Examples of trajectories obtained from the field model \eqref{sine-gordon} corresponding to unstable regions, marked as light red areas in Figure \ref{fig_B07}, are presented in the left panels of Figure \ref{fig_B08}, i.e., panels (a) and (c). 
These trajectories illustrate the escape process from the potential minimum located at $x_s = -6$. In both panels, the forcing frequency is set to $\omega = 0.03$, the wavevector is $k = \pi/6$, and the parameter $\Omega$ is fixed at $0.04$. The black solid lines represent the results obtained from the full PDE model, whereas the orange dashed lines (shown for comparison) correspond to the reduced effective ODE model. Panel (a) corresponds to $\varepsilon_1 = 0.05$, while panel (c) shows the case with $\varepsilon_1 = 0.07$. In both scenarios, the initial kink position is aligned with the minimum at $x_0(0) = -6$. 
The background colour map in each panel indicates the values of the function $\mathcal{F}$, as specified in the accompanying legend. 

On the other hand, the right panels (b) and (d) of Figure \ref{fig_B08} show example trajectories (obtained from the field model \eqref{sine-gordon}) corresponding to stable areas, i.e., the blue areas in Figure \ref{fig_B07}. These trajectories illustrate oscillations around the minimum at $x_s=-6$. In the figure, the forcing frequency $\omega$ is set to $0.03$, the wavevector is equal to $k=\pi/6$, and the $\Omega$ parameter is $0.05$ at panel (b) while $\Omega=0.1$ at panel (d). The values of parameter $\varepsilon_1$ are $0.02$ at panel (b) and $0.07$ at panel (d). The trajectory obtained based on the field model is represented as a solid black line. For comparison, the dashed orange line represents  the result of the effective model \eqref{1dof_ansatz}. In all these cases, the kink initially rests in the position corresponding to the minimum of $\mathcal{F}$ function, i.e., $x_0(0)=-6$.
Despite the observed oscillations, the stable evolution of the
dynamics results in the bounded motion of the kink within the
well in which it was originally. 
The trajectories presented in panels (a)–(d) on the 
$(\varepsilon_1, \Omega)$ plane
correspond to the points marked (a)-(d) in Figure \ref{fig_B07} (b).

\FloatBarrier

\section{Conclusions and Future Challenges}
In this study, we investigated how different types of 
spatio-temporal inhomogeneities can affect the dynamics of kinks in a perturbed sine-Gordon model. A distinctive feature of our approach is that the system is explicitly non-autonomous: time dependence enters directly into the coefficients of the governing equation. This temporal modulation opens up new possibilities, such as controlled transport of the kink across a periodic (or,
more generally, spatially heterogeneous) potential landscape. To capture and interpret these effects, we construct an effective model based on a suitably tailored ansatz that explicitly incorporates the influence of spatio-temporal inhomogeneities. As is common in this context \cite{CKW14}, the ansatz reduces the dynamics to a single collective coordinate that tracks the position of the kink’s center — defined as the spatial point where the scalar field attains the value $\pi$.

Our analysis focuses on the dissipationless regime, where energy loss mechanisms are absent. {In the case of dissipation, the analysis can be performed based on non-conservative Lagrangian method \cite{Dobrowolski2025}.}
Within this setting, the proposed effective model proves to be reasonably accurate: its reduced dynamics, governed by ordinary differential equations, closely mirror the full trajectories derived from the original field-theoretic formulation. Remarkably, this correspondence persists over considerable time intervals — extending across tens or even hundreds of characteristic time units. The predictive strength of the model becomes particularly evident in the presence of non-autonomous effects, where the parameters of the governing equation vary explicitly with time and generate complex, nontrivial kink dynamics. These temporal variations are prototypically modelled herein as traveling modulations — waves propagating through the medium — which, even in the absence of external driving forces, can induce directed motion of the kink. Such spontaneous transport, especially when it results in transitions between distinct equilibrium configurations, is of particular interest. It offers a pathway towards controlling the position of solitary waves (at the level of our
effective description), a subject actively explored in diverse settings ranging from discrete lattices \cite{hector}, through heterogeneous continuum systems \cite{theocharis}, to media with balanced gain and loss, both in theory \cite{Rossi2024} and in experimental realizations \cite{Jang2015}.

We further carried out a stability analysis using the Floquet framework. Within the linearized regime, this led to a Mathieu-type equation, enabling us to delineate the boundaries of potential parametric instability regions in the relevant  parameter space — characterized by the interplay between the amplitude and 
frequency of the perturbation. The resulting stability diagram was then compared to that obtained from direct simulations of the full field model. Interestingly, both approaches yield qualitatively similar phenomenology regions, especially in the upper portions 
of
the so-called Arnold tongues. In contrast, the lower regions reveal more intricate behaviour in the field model, where alternating intervals of stability and instability emerge — the kink either remains trapped in oscillatory motion or escapes from it, depending sensitively on the parameter values. Notably, in the stable zones, the agreement between the effective approximation and the full field dynamics is essentially exact.

The present study naturally paves the way towards several promising research directions. One particularly compelling avenue involves the analysis of effective models with two degrees of freedom, which allow for a more refined description of kink dynamics, including both translational and internal (e.g., width-related) modes, which are known to potentially emerge in the presence of perturbations~\cite{CKW14}. Investigating the stability of such systems—especially in the presence of time-dependent parametric perturbations—offers a fertile ground for understanding phenomena such as parametric resonance, mode coupling, and the potential onset of chaotic behaviour, especially given the non-autonomous
nature of the relevant drive (e.g., in a way reminiscent of the
forced van der Pol oscillator or similar, effectively 3-dimensional
dynamical systems). These effective models provide a computationally efficient and conceptually transparent framework in which one can explore the conditions for stable motion, controlled transport, and resonance-induced instabilities, all of which are relevant before transitioning to the full field-theoretic setting.
Parallel lines of inquiry concern the role of non-autonomous perturbations and their potentially chaotic dynamics, as well 
as the higher-dimensional versions of the sine-Gordon model. The extension to two or more spatial dimensions introduces new dynamical features and structural robustness of kinks, and the interplay between spatial complexity, transverse dynamics and temporal modulation remains largely unexplored. In particular, time-dependent coefficients in such multidimensional settings may lead to novel mechanisms of wave guidance, spontaneous motion, or resonance-driven instabilities, with potential relevance in physical contexts such as large area Josephson junctions with spatially varying surface properties. These topics are currently being pursued and will be addressed in forthcoming work.

\section{Appendix}
The effective Lagrangian \eqref{Leff} involves parameters explicitly computed via the integrals listed below
\begin{equation}
\begin{gathered}
\label{integrals}
    M = 4\int_{-\infty}^{+\infty}dx\sech^2(\xi) \, W^2(\xi),\\
\alpha = 2\int_{-\infty}^{+\infty}dx\sech^2(\xi) \, \frac{f(t,x_0)}{g(t,x_0)} \, W(\xi) \xi
    , \\
    V = 2\int_{-\infty}^{+\infty}dx\sech^2(\xi) \, \left( g(t,x) + g(t,x_0) \, \frac{\mathcal{F}(x)}{\mathcal{F}(x_0)}\gamma^2 - \frac{f^2(t,x_0)}{4 g^2(t,x_0)} \, \xi^2 \right) .
\end{gathered}
\end{equation}
The auxiliary functions $W$ and $f$ involved in the coefficient expressions are given below
\begin{equation} \label{W}
    W(\xi) = \frac{1}{2} \, \left( \frac{\partial_{x_0} {\cal F}(x_0)}{ {\cal F}(x_0)} - \frac{\partial_{x_0} { g}(t,x_0)}{ {g}(t,x_0)} \right) \, \xi + \sqrt{\frac{g(t,x_0)}{{\cal F}(x_0)}} \,  , \,\,\,\,\, f(t,x_0) = \omega \varepsilon_1 \cos{(k x_0 - \omega t)} .
\end{equation}
All coefficients depend on the dynamic variable $x_0(t)$, and additionally exhibit explicit time dependence. Most of the integrals mentioned above can be evaluated by expressing the coefficients explicitly in terms of the functions ${\cal F} (x_0)$ and $g(t, x_0)$, which appear in the Lagrangian density
\begin{equation}
\begin{gathered}
\label{integrals-fin}
    M = \frac{\pi^2}{6 } \sqrt{\frac{{\cal F}(x_0)}{g(t,x_0)}} \left( \frac{\partial_{x_0}{\cal F}(x_0)}{{\cal F}(x_0)} - \frac{\partial_{x_0} g(t,x_0)}{g(t,x_0)}\right)^2 + 8 \sqrt{\frac{g(t,x_0)}{{\cal F}(x_0)}},\\
\alpha = \frac{\pi^2}{6 } \, \sqrt{\frac{{\cal F}(x_0)}{g(t,x_0)}} \left( \frac{\partial_{x_0}{\cal F}(x_0)}{{\cal F}(x_0)} - \frac{\partial_{x_0} g(t,x_0)}{g(t,x_0)}\right) \frac{f(t,x_0)}{g(t,x_0)}  , \\
    V = 2 \int_{-\infty}^{+\infty} dx \sech^2 \xi \, g(t,x) +\frac{2  g(t,x_0)}{{\cal F}(x_0)} \int_{-\infty}^{+\infty} dx \sech^2 \xi \, {\cal F}(x) - \frac{\pi^2 f^2(t,x_0)}{12  g^2(t,x_0)} \sqrt{\frac{{\cal F}(x_0)}{g(t,x_0)}}  .
\end{gathered}
\end{equation}

\section{Acknowledgments}
This material is based upon work supported by the U.S. National Science Foundation under the awards PHY-2110030, DMS-2204702 and PHY-2408988 (PGK). Research project supported by program "Excellence Initiative - Research University" for the AGH University of Krakow (JG).

\FloatBarrier
\printbibliography

\end{document}